\documentclass[preprint,12pt,number]{elsarticle}

\usepackage{amssymb,booktabs}
\usepackage{amsmath}
\usepackage{url}
\usepackage{natbib}

\journal{Physica A}

\begin{document}

\begin{frontmatter}



\title{Gap distributions between successive personal bests in cricket: Data and Models} 


\author[label1]{Priyanka D. Bhoyar} 
\author[label2]{Prashant M. Gade}
\address[label1]{Department of Physics, Seth Kesarimal Porwal College,
Rashtrasant Tukadoji Maharaj Nagpur University,
Kamptee 441001, Maharashtra, India}

\address[label2]{Department of Physics,
Ramniranjan Jhunjhunwala College of Arts, Science and Commerce,
Ghatkopar West, Mumbai 400086, Maharashtra, India}

\begin{abstract}
Successive personal best performances provide a natural measure of progression in an athlete's career. Classical record theory predicts a universal gap distribution, $P(g)\sim 1/g$, for independent and identically distributed (i.i.d.) sequences. However, sporting careers are shaped by learning, aging, changes in ability, and external influences that violate these assumptions.
We investigate the statistics of inter-record gaps, defined as the number of innings between successive personal best scores, in cricket. Using career records of leading Test, ODI, and T20 players obtained from ESPN Cricinfo.
We find that the empirical distributions are well described by truncated power law $P(g) \propto g^{-\alpha} e^{-\lambda g}$ with exponents in the range (0.799 $\leq \alpha \leq$ 0.843). Much of this deviation disappears when the temporal ordering of innings is destroyed, indicating that career evolution plays a key role in shaping record occurrence. Bootstrap-shuffled careers, which preserve individual score distributions and career lengths while removing temporal ordering, yield significantly larger exponents ($\alpha \approx 0.939\text{--}0.979$).
These findings show that the progression of personal best performances retains information about the temporal organization of a player's career and cannot be fully explained by simple stochastic record processes. More generally, they illustrate how record statistics are altered in nonstationary and path-dependent systems.
\end{abstract}

\begin{graphicalabstract}
\includegraphics[width=1.2\linewidth]{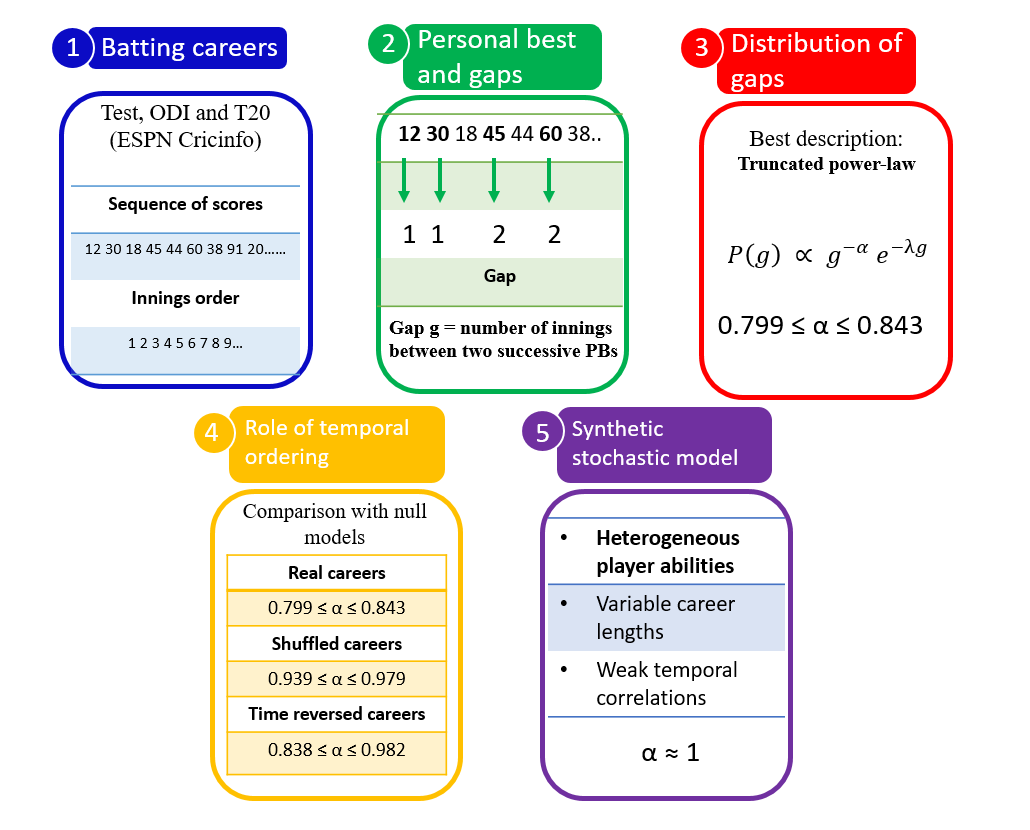}
\end{graphicalabstract}

\begin{highlights}
\item personal best gap statistics provide a quantitative measure of career progression.
\item Inter-record gap distributions in cricket exhibit truncated power law scaling.
\item Maximum-likelihood analysis identifies the truncated power law as the preferred model.
\item Shuffled careers demonstrate that temporal ordering strongly influences record statistics.
\item Synthetic null models explain broad trends but not all empirical scaling behavior.
\end{highlights}


\end{frontmatter}


\section{Introduction} 
 In older times, contests emphasized skill rather than measurable performance. 
 Modern times have facilitated measurements and records are retained forever. 
 Performance progress has become central theme in sports \cite{boccia2021performance}. Athletes, coaches, and fans often evaluate careers through milestones such as centuries scored, wickets taken, and personal best performances. Cricket occupies a distinctive place in this regard because of its extensive statistical culture that often celebrates highly specific personal milestones and records along with team outcome.
Detailed scorecards are available for virtually every professional match, allowing careers to be reconstructed innings by innings and making cricket an attractive setting for studying performance progression.

Comparisons based on absolute records are common, but they are often complicated by differences in era, playing conditions, equipment, rules, and opportunities available to players \cite{gembris2007evolution}. Absolute numbers rarely tell the whole story. Players with higher batting averages than Viv Richards exist, yet such comparisons alone do not necessarily imply superior ability because performances are shaped by era and playing conditions. Personal bests provide a useful alternative perspective because they measure improvement relative to a player's own past achievements and therefore capture performance progression within a career \cite{stevenson2018modelling}. The occurrence of a new personal best represents an important milestone, while the intervals between successive personal bests quantify how frequently a player surpasses previous achievements. As such, the gap between successive personal bests provides a natural measure of career progression, with short gaps corresponding to rapid improvement and longer gaps indicating extended periods without a new milestone. Yet career progression, aging, and the competitive nature of sport do not permit relentless improvement \cite{berthelot2015has}. As performance approaches an individual's limits, achieving a new personal best becomes increasingly difficult, making the timing of successive personal bests an interesting measure of performance evolution.

The three formats in cricket differ substantially in match duration, scoring patterns, and tactical constraints, providing an opportunity to examine whether record progression depends on the competitive environment. In Test cricket, players often have greater opportunities to build long innings and pursue personal milestones, whereas in T20 cricket strategic constraints and time pressure limit such opportunities. Fans and commentators frequently discuss periods of exceptional form, prolonged slumps, and career defining performances, raising the question of whether such patterns reflect genuine structure in performance or arise naturally from random variation \cite{ram2022significant}. Motivated by these considerations, we investigate the temporal patterns of personal best scores of top players across Test, ODI, and T20 cricket careers through the distribution of gaps between successive personal bests.

Statisticians have studied record processes extensively since the 1950s, and classical record theory provides a general framework for analysing record breaking events in sequences of independent and identically distributed observations \cite{arnoldrecords,krugrecords,Wergenrecords}. These results establish baseline predictions for the occurrence of new records and therefore provide a natural null model for sporting performances. 
Such comparisons test whether a player's scores are consistent with random draws from a fixed ability distribution or whether they contain additional temporal structure arising from correlations, memory effects, or evolving performance.
Cricket performance has previously been shown to exhibit long-range correlations, bursty dynamics, and other non-trivial temporal features \cite{ribeirocricket,burstcricket}, consistent with similar heavy-tailed patterns observed in many human activities \cite{barabasibursts}. Despite these advances, the statistical properties of successive personal best achievements have received little attention. Here we investigate the distribution of gaps between successive personal best scores across cricket careers. 
To determine whether the observed gap statistics arise from simple stochastic mechanisms or from the temporal structure of batting careers, we compare the empirical data with two null models: randomly shuffled careers, which preserve each player's score distribution while destroying temporal correlations, and time-reversed careers. We further compare the results with synthetic stochastic models incorporating heterogeneous player abilities and career lengths to assess whether these ingredients alone are sufficient to reproduce the observed statistics.

\section{Theory}
\subsection{Probability of setting a record in the $n^{th}$ Inning}

It is well-known that, under simple stochastic assumptions, the probability that a player sets a personal best in the $n^{th}$ inning decreases roughly as $1/n$ \cite{arnoldrecords,Wergenrecords}. The reasoning is straightforward. If all $n$ scores are independent and identically distributed (i.i.d.), each of the $n$ scores is equally likely to be the maximum among the first $n$ innings. Therefore, the probability that the latest score is a new personal record is
$
R(n) = \frac{1}{n}
$. This is expected because it becomes more difficult to break the record 
as one progresses in career.

We now give a heuristic argument for  the distribution of gaps between personal bests. Formally, for a player with scores $(S_1, S_2, \dots, S_N)$, 
a \emph{personal best} occurs at innings $n$ if
$S_n > \max(S_1, \dots, S_{n-1})$.
We define the \emph{gap} (G) as the number of innings between successive personal bests. To study the gap distribution, we compute the gaps for individual players and then aggregate across all players to obtain the overall frequency distribution.

As mentioned above,
\begin{equation}
R(n) = \mathbb{P}(\text{record at } n) = \frac{1}{n}
\end{equation}
Consider the gap $G$ between successive records. 
Conditional on the previous record occurring at inning $n$,
the probability that the next record occurs after a gap of $g$ 
innings is

\begin{equation}
 \mathbb{P}(G=g \mid n) = R(n+g) \prod_{k=1}^{g-1} \bigl(1 - R(n+k)\bigr)
= \frac{1}{n+g} \prod_{k=1}^{g-1} \left(1 - \frac{1}{n+k}\right).   
\end{equation}

\noindent
Now
\begin{equation}
\prod_{k=1}^{g-1} \left(1 - \frac{1}{n+k}\right) = \frac{n}{n+g-1,},
\end{equation}
Thus 
\begin{equation}
\mathbb{P}(G=g \mid n) = \frac{1}{n+g} \cdot \frac{n}{n+g-1} \approx \frac{n}{(n+g)^2}.
\end{equation}

\noindent
To obtain the \emph{unconditional} gap distribution, we average over all possible previous record positions $n$:

\begin{equation}
P(g) = \sum_{n} \mathbb{P}(G=g \mid n). R(n)
 \approx \sum_{n=1}^{\infty} \frac{1}{n} \cdot \frac{n}{(n+g)^2} = \sum_{n=1}^{\infty} \frac{1}{(n+g)^2}.
\end{equation}

\noindent
Replacing the sum by an integral for large (g), it can be easily seen that

\begin{equation}
P(g) \approx \int_{g}^{\infty} \frac{dx}{x^2} = \frac{1}{g}.
\end{equation}

\noindent
Thus, the gap distribution is expected to be 
$
P(g) \sim \frac{1}{g}.
$
Batting performances in cricket, however, are not expected to satisfy the assumptions of the i.i.d. model. A player's ability evolves through experience, aging, injuries, changes in technique, and fluctuations in form. These factors introduce temporal correlations and non-stationarity into career trajectories, causing successive innings to be statistically dependent. We also note that the classical record-statistics prediction assumes that new record values remain available throughout the observation period. For variables with a very small set of attainable values, record-breaking opportunities can be rapidly exhausted, leading to strong finite-size effects and departures from the asymptotic behavior. For example, if a variable can take only the values 0 and 1, the first occurrence of 1 is the last possible record, and the corresponding gap distribution is geometric rather than scale free.

\section{Data and Modeling Framework} 
\subsection{Empirical Data}	
We began by extracting data from the ESPN Cricinfo pages listing the leading run scorers in each format (Test, ODI, and T20) and iteratively navigated to each player's career page using a python-based scraping pipeline\cite{cricinfostatsguru}. The pipeline extracted all innings scores in chronological order, automatically filtering out innings marked as Did Not Bat (DNB) or Team Did Not Bat (TDNB). Using these scores, we computed the gaps between successive personal bests for each player.

The classical prediction for record statistics is an asymptotic result that is expected to emerge only when sufficiently long sequences are available. Therefore, two competing considerations arise in constructing the dataset. On the one hand, players must have sufficiently long careers to generate meaningful record-gap statistics and permit comparison with the classical prediction. On the other hand, a sufficiently large pool of players is required to obtain reliable aggregate gap distributions.

For Test and ODI cricket, all players appearing in the corresponding ESPN Cricinfo career-runs rankings were included, resulting in datasets of 108 and 98 players, respectively. For T20 cricket, the available ranking contained substantially more players, many of whom had relatively short careers. Since our analysis fits a truncated power law distribution, players with very short careers impose a smaller finite-size cutoff on the observable gap distribution, reducing the range over which scaling can be assessed and potentially biasing the estimated exponent toward larger values. The T20 dataset was therefore restricted to the top 101 run scorers.

The resulting datasets have average career lengths of 183, 227, and 91 innings for Test, ODI, and T20 cricket, respectively. The relatively small number of innings in T20 cricket reflects the youth of the format. Indeed, even the leading T20 run scorer, Babar Azam, had played only about 136 innings at the time of data collection. Consequently, there is currently no straightforward way to construct a T20 dataset with career lengths comparable to those available in Test and ODI cricket.

Our automated approach ensures reproducibility and allows rapid expansion to larger datasets or future seasons without manual intervention.

\subsection{Models Tested}

\textbf{A) Empirical Data:} We analyze innings-by-innings records for individual batsmen. For each player, we identify the innings in which a new personal best score was achieved, i.e., an innings score exceeding all previous scores in that player's career. We then compute the gaps between successive personal best innings and aggregate these gaps across all players to obtain the overall gap distribution. 
Significant deviations in the values of the exponent of truncated power law ($\alpha$) appear in the tail of the distribution. These departures indicate that real batting performance does not fully satisfy the assumptions of a simple i.i.d. process.

\textbf{B) Bootstrap Shuffle:} To isolate the effect of temporal ordering, we randomly shuffle the innings scores of each player. This procedure preserves the set of scores and hence the overall scoring distribution of each batter, while removing any temporal correlations present in the original sequence.

\textbf{C) Reversed Careers:} As an additional test, we reverse the chronological order of innings for each player and recompute the gap distribution. This transformation preserves both the set of scores and their statistical distribution, while altering the temporal structure of record occurrences.

\textbf{D) Synthetic Models:} To investigate the origin of the observed deviations from the theoretical prediction, we compare the empirical results with several synthetic datasets. We generate careers for 100 players, each consisting of 400 innings, under the following scenarios:

\begin{enumerate}
\item Homogeneous careers with i.i.d. scores drawn from a common distribution.
\item Heterogeneous players, where each player is assigned a distinct scoring distribution (e.g., different mean or variance).
\item Heterogeneous career lengths. Here, we do not have same 400 innings for every player.
\item Careers with temporal correlations, including both weak short-range correlations and long-range power law correlations in scores.
\end{enumerate}

The first three models produce results qualitatively similar to the bootstrap-shuffled data and do not reproduce the deviations observed in the empirical records. Introducing temporal correlations alters the gap distribution, but the resulting behavior remains inconsistent with that found in the real batting data. This indicates that these models are inadequate for explaining the observed behavior and that additional temporal structure beyond above factors is required.

\section{Results and Analysis}
\subsection{Probability of setting record in $n^{th}$ innings}
\begin{figure}[h!]
    \centering
    \includegraphics[width=\textwidth]{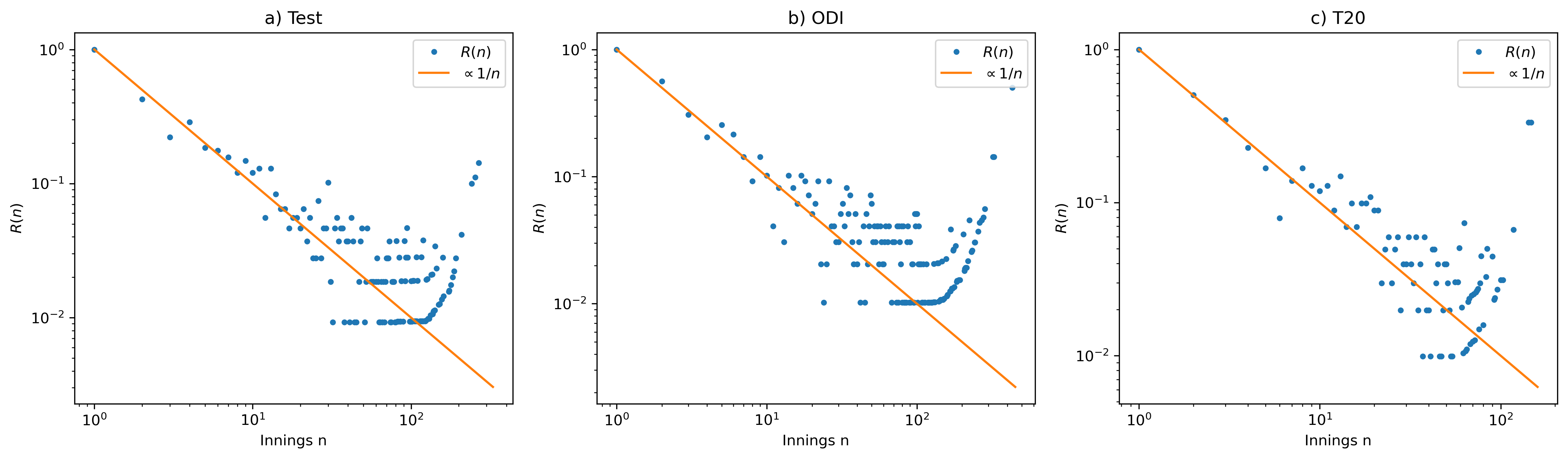}
    \caption{Shows the plot of probability $R(n)$ of setting a new record in the $n^{th}$ inning on log-log scale for (a) Test Matches (b) ODI Matches and (c) T20 Matches. Early innings follow an approximate $1/n$ behavior, while deviations appear in the tail due to heterogeneity in career lengths.}
    \label{fig:R_n}
\end{figure}

To quantify this for real cricket data, we define
\[
R(n) = \frac{\text{Number of players setting a new record in their $n$-th inning}}{\text{Number of players who played at least $n$ innings}}.
\]
Fig.~\ref{fig:R_n} shows the probability R(n) of setting a new record at the $n^{th}$ innings for (a)Test, (b)ODI, and (c)T20 formats.  The higher
values for large $n$ come from heterogeneous career lengths.
For instance, Sachin Tendulkar set a record in his $431^{st}$ innings, at that point, only two players in the dataset had career lengths exceeding $431$ innings, giving $R(431) \simeq 0.5$. This illustrates that heterogeneity in career lengths strongly affects the probability of setting late-career records.

\subsection{Normalized probability mass function(PMF)}
\begin{figure}[h!]
    \centering
    \includegraphics[width=\textwidth]{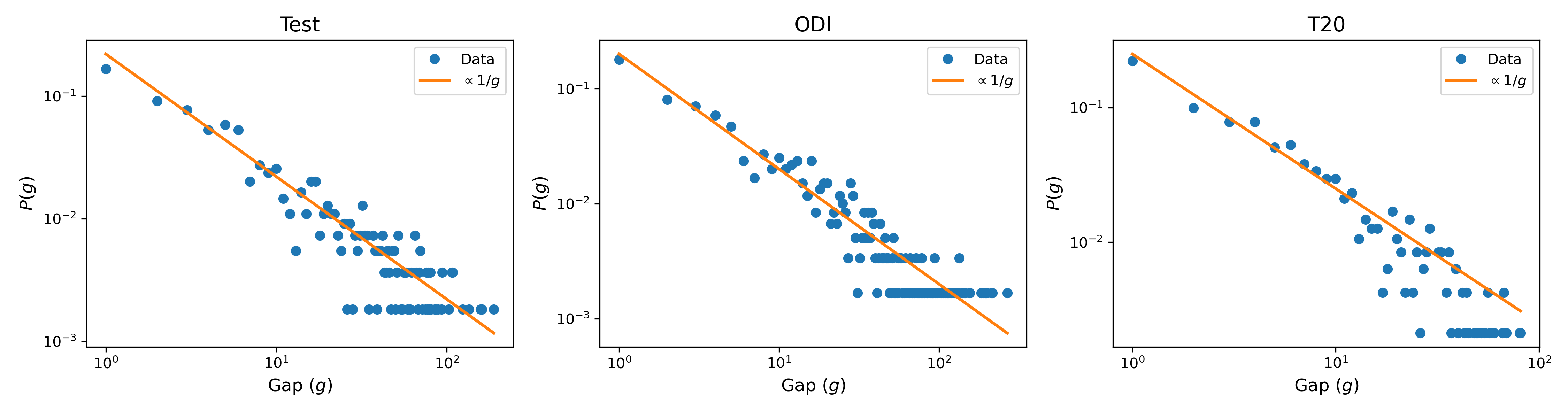}
    \caption{ Normalized probability Mass function (PMF) of the gap distribution for the empirical Test, ODI, and  T20 matches. The distribution shows heavy-tailed behavior and deviations from the classical 1/g record 
    statistics prediction.}
    \label{fig:pdf_test}
\end{figure}

We now turn to the distribution of gaps between personal bests. The empirical gap distributions for  Tests, ODIs, and T20s data are first shown using their normalized  Probability mass function (PMF). 
\noindent
A probability mass function $P(g)$ is probability that the gap between successive records is exactly g. 
\begin{equation}
P(g) = \mathbb{P}(Gap = g), 
\end{equation}
$P(g)$ is estimated from the empirical data as
\begin{equation}
P(g) = \frac{N(g)}{\sum_g N(g)},
\end{equation}
where $N(g)$ denotes the number of occurrences of gap $g$.
The normalized gap distributions P(g) for Test, ODI, and T20 matches are shown on log–log scales in Fig.~\ref{fig:pdf_test}.
All formats display heavy-tailed behavior. 
Although, the data approximately follow a power law decay, comparison with the 1/g reference line reveals noticeable deviations, indicating a slower decay and supporting the presence of truncated power law scaling.

\subsection{Truncated Power law Fitting}
While the gap distributions exhibit approximate power law behavior at small scales, clear deviations are observed at larger gaps (see Fig.\ref{fig:pdf_test}). 
These deviations arise due to finite-size effects, heterogeneity, and correlations in the system. To account for this behavior the gap distribution $P(g)$ was fitted to a truncated power law for each data set using maximum likelihood estimation. 
\begin{equation}
    P(g) \propto g^{-\alpha} e^{-\lambda g}
\end{equation}

\begin{equation}
P(g)=\frac{g^{-\alpha}e^{-\lambda g}}
{Z(\alpha,\lambda)},
\end{equation}
where $\lambda$ is the cuttoff parameter and Z is the normalization constant,

\begin{equation}
Z(\alpha,\lambda)=
\sum_{g=g_{\min}}^{g_{\max}}
g^{-\alpha}e^{-\lambda g}
\end{equation}


All PMF datasets exhibit an approximately linear trend on log-log axes, with a noticeable downward curvature at large gap values, indicating a deviation from pure power law behavior. However, the tail regions are sparse and subject to significant fluctuations, making direct inference from the probability mass function (PMF) less clear. To address this, we visualize using the complementary cumulative distribution function (CCDF)

\subsubsection{Complementary cumulative distribution function(CCDF)}
The complementary cumulative distribution function (CCDF) is defined as:
\begin{equation}
R(g)=P(G\ge g),
\end{equation}
where G denotes the inter-record gap. Equivalently,

\begin{equation}
R(g)=\sum_{g' \ge g} P(g'),
\end{equation}
where P(g) is the probability mass function of the gap distribution. Compared to the PMF, the CCDF provides a smoother representation of the tail. The CCDF is shown for visualization only. Model parameters were estimated by maximum-likelihood fitting of the observed gap distribution, with the negative log-likelihood minimized using \textit{scipy.optimize.minimize}. 
The fitted parameters were subsequently used to generate the model CCDF. The plots show an excellent fit (See Fig.\ref{fig:ccdf}).

\subsection{Null Models Based on Career Reordering}
The empirical gap distributions differ substantially from the
classical prediction of record statistics. To identify the origin
of this discrepancy, we construct two null models directly from
the empirical data. Unlike synthetic score-generation models,
these procedures preserve the actual scores achieved by every
player and therefore retain the empirical distributions of
batting performance. As mentined above, we also construct synthetic models. The gap distributions obtained from the original, shuffled, 
reversed careers and synthetic models are analyzed using identical procedures and are
fitted using the truncated power law form.

\subsubsection{Bootstrap-Shuffled Careers}

This procedure preserves a) the career length of each player,
b) the batting average,
c) the full score distribution,
d) the frequency of high and e) low scores.
However, all temporal ordering and correlations between innings
are destroyed. With this change the fitted exponents $\alpha$ increase 
to 0.939-0.979 indicating that the temporal correlations play an important role.

\subsubsection{Reversed Careers}

Reversing a career preserves the sequence of scores and hence retains correlations, but reverses the direction of career progression. This modification also increases the fitted exponent $\alpha$. The result suggests that improvements leading to new personal bests are more likely to occur later in a player's career than would be expected from a temporally symmetric process.


\subsubsection{Synthetic Models}

The synthetic models described in the previous section were analyzed using the same methodology as the empirical and reordered careers. The resulting exponents remained close to the classical record-statistics prediction ($\alpha \approx 1$), even for models incorporating heterogeneous player abilities, variable career lengths, and weak temporal correlations. Since none of the synthetic models reproduced the substantially smaller exponents observed in the empirical data, they do not provide additional explanatory power beyond the bootstrap and career-reordering analyses. We therefore restrict further discussion to the empirical and null-model results.

\section{Effect of Temporal Ordering}

\begin{figure}[h!]
\centering
\includegraphics[width=\textwidth]{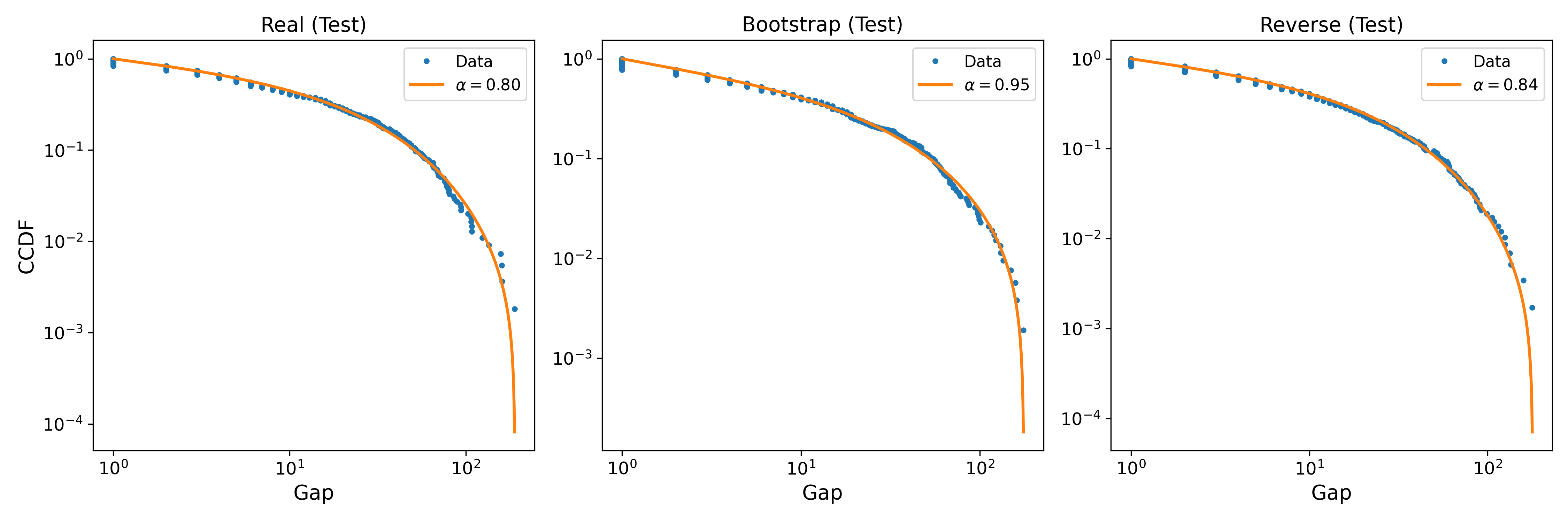}
\includegraphics[width=\textwidth]{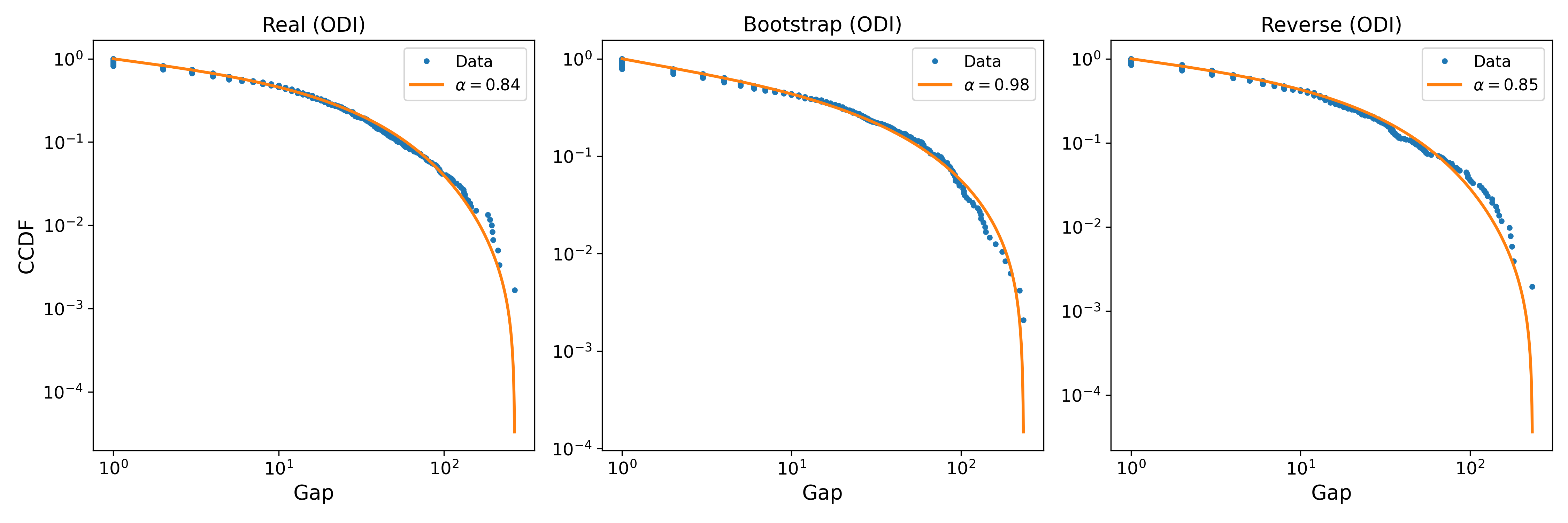}
\includegraphics[width=\textwidth]{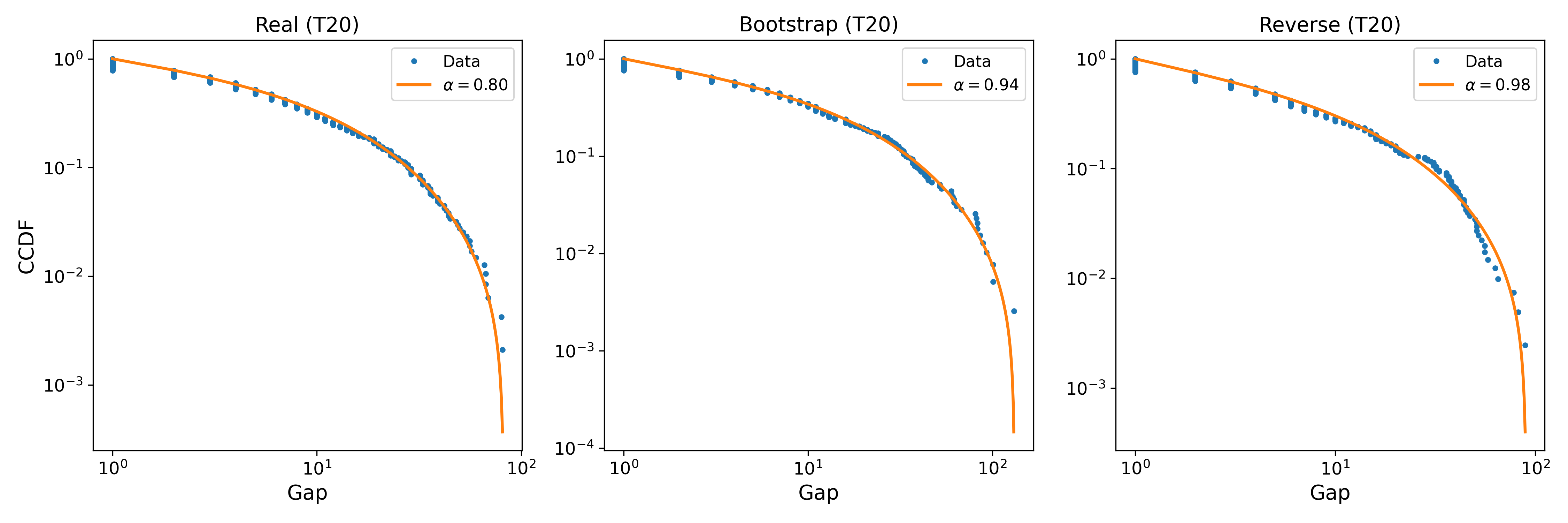}
\caption{Complementary cumulative distribution functions (CCDFs) of inter-record gaps for (a) Test, (b) ODI, and (c) T20 cricket careers. The distributions exhibit broad tails spanning several orders of magnitude, indicating substantial variability in the waiting times between successive personal best performances. The solid lines represent maximum-likelihood fits to the truncated power law form $P(g)\sim g^{-\alpha}e^{-\lambda g}$. The estimated parameters are $\alpha=0.799,~\lambda=0.0180$ (Test), $\alpha=0.843,~\lambda=0.01447$ (ODI), and $\alpha=0.799,~\lambda=0.035$ (T20) for real data. 
For bootstrap-shuffle the parameters are $\alpha=0.952,~\lambda=0.008$ for test, $\alpha=0.979,~\lambda=0.008$ for ODI and $\alpha=0.939,~\lambda=0.021$ for T20.
For reverse data, the parameters are $\alpha=.838,~\lambda=0.020$ for test, 
$\alpha=0.854,~\lambda=0.016$ for ODI and $\alpha=.982,~\lambda=0.023$ for T20. 
}
    \label{fig:ccdf}
\end{figure}

Fig.~\ref{fig:ccdf} compares the inter-record gap
distributions obtained from the original careers, the
bootstrap-shuffled careers, and the reversed careers. 

The bootstrap-shuffled data exhibit a markedly different
behavior from the empirical records. In all three formats, the
gap distributions are consistent with the classical prediction
of record statistics,
$P(g)\sim \frac{1}{g}$,
with fitted exponents close to unity. Since the bootstrap
procedure preserves career lengths, batting averages, and the
complete score distribution of every player, this result
demonstrates that these factors alone cannot explain the
empirical scaling observed in cricket.
  In contrast, the empirical datasets consistently yield smaller
exponents, $\alpha \approx 0.799 - 0.843$, indicating substantially broader waiting-time distributions
between successive personal best performances.

Interestingly, reversing the order of innings does not restore
the classical exponent for tests and ODI. The reversed careers continue to
display broad-tailed gap distributions with exponents
significantly below unity. Thus, the observed deviation from
classical record statistics cannot be attributed solely to
monotonic improvement or decline during a player's career.
T20 is the latest format and most of the players we studied are still playing and their career is far from over.

One intuitive mechanism is the presence of a "record ceiling." An exceptionally high score achieved during a career can become difficult to surpass and may effectively terminate the sequence of future personal records. Famous examples include Garfield Sobers' 365*, scored at the age of 21, and Karun Nair's unbeaten 303 in only his third Test innings. In both cases, the players never exceeded these scores despite substantial subsequent careers. Such extreme performances illustrate how unusually large personal bests can strongly influence record dynamics. However, they do not fully explain the observed scaling. The comparison between original and time-reversed careers suggests that the temporal structure of major records also matters. Reversed careers yield systematically larger exponents than the original data, indicating that record setting performances tend, on average, to occur later in real careers than expected under time reversal. Thus, while extreme scores such as those of Sobers and Nair provide striking examples of record ceilings, the observed heavy tails likely reflect a broader pattern of career evolution and late-career performance peaks.

Smaller exponents correspond to fatter tails in the gap distribution. Consequently, the probability of surpassing a personal best even after a long interval remains significantly higher in real cricket careers than would be expected for i.i.d. random performances. This persistence occurs despite the existence of an upper performance ceiling. One possible explanation is the gradual improvement of player skills and experience over the course of a career. In addition, changes in playing conditions and rules over time have generally favored higher scoring. There has been improvement in protective equipment, bat technology and training methods. All these factors provide further opportunities for new personal records.

Test matches should be distinguished other two formats. Test matches permit much longer innings, allowing batters to continue accumulating runs and reach new milestones. Furthermore, the possibility of a draw, in contrast to the strictly win–loss outcomes of limited-overs formats, reduces the pressure for rapid scoring and can extend opportunities for exceptional individual performances.

For T20 cricket, caution is warranted when interpreting these results. Even among the top 101 run-scorers considered here, average career lengths are substantially shorter than in the longer formats of the game.  As the format matures and players accumulate substantially longer careers, it will be interesting to revisit these record-gap statistics in the coming decade to determine whether the observed scaling behavior persists.

These observations identify temporal ordering as the primary
source of the observed scaling behavior. Since the bootstrap-shuffle removes only the ordering of performances while
preserving all other statistical properties of the data, the
difference between the real and shuffled careers provides
direct evidence that cricket careers contain non-trivial temporal
structure beyond that expected from independent random
performances.


We would like to mention that distributions other than truncated power law fit were considered. We fitted the gap distributions using Poisson, lognormal, negative binomial, and truncated power law models using maximum likelihood estimation and Akaike Information Criterion. In all three formats, the truncated power law provided the lowest AIC by a substantial margin, indicating that it offers the best description among the candidate distributions considered. 
The fitted truncated power law parameters are summarized in Table.~\ref{table2} For the empirical datasets, the scaling exponents range from $(\alpha=0.799-0.843)$. The corresponding cutoff parameters indicate finite-size effects arising from limited career lengths and observation windows. Overall, the log likelihood and AIC-based comparisons favor the truncated power law description, suggesting that inter-record gaps are characterized by broad, heavy-tailed statistics spanning multiple temporal scales. Error bars on the truncated power law exponent were estimated using the profile-likelihood method. For each fixed value of $\alpha$, the likelihood was maximized with respect to the cutoff parameter $\lambda$. The $1\sigma$ confidence interval was then obtained from the range of $\alpha$ values satisfying
$
2\left[\log L_{\max}-\log L(\alpha)\right]\le 1,
$
where $L_{\max}$ is the maximum likelihood and $L(\alpha)$ is the profile likelihood.

\begin{table}[h!]
\centering
\caption{Truncated power law parameters for real, bootstrap-shuffled and reversed career.}
\begin{tabular}{lcccc}
\toprule
\hline
Dataset & $\alpha$ & $\lambda$ & Log-likelihood & AIC \\
\midrule
\hline
Test & 0.799 $\pm $ 0.05 & 0.0180 & -2029.09 & 4062.15 \\
ODI & 0.843 $\pm$ 0.045 & 0.01447 & -2266.91 & 4537.82 \\
T20 & 0.799 $\pm$ 0.07 & 0.0357 & -1494.60  & 2993.21 \\
\hline
Test (Bootstrap) & 0.952 $\pm$ 0.05 & 0.008 & -1709.46 & 3422.93\\
ODI (Bootstrap) & 0.979 $\pm$ 0.045 & 0.009 & -1807.22 & 3618.44\\
T20 (Bootstrap) & 0.939 $\pm$ 0.065 & 0.021 & -1276.06 & 2556.12\\
\hline
Test (Reverse)& 0.838 $\pm$ 0.05 & 0.020& -2072.9& 4149.80\\
ODI (Reverse) & 0.854 $\pm$ 0.05 & 0.016 & -1876.91 & 3757.81\\
T20 (Reverse) & 0.982 $\pm$ 0.07 & 0.023 & -1247.22 & 2498.44\\
\hline
\bottomrule
\end{tabular}
\label{table2}
\end{table}

\section{Discussion}

A central result of this study is the systematic departure of real cricket careers from classical record statistics predictions. For an i.i.d. sequence of performances, the probability that the $n^{\mathrm{th}}$ observation is a new record is $1/n$, implying an inter-record gap distribution $P(g)\sim 1/g$ and therefore a scaling exponent $\alpha \approx 1$ \cite{arnoldrecords}.

To test whether the observed deviation could be explained by player heterogeneity or career length, we constructed bootstrap-shuffled careers in which the innings of each player were randomly reordered. This procedure preserves career lengths and score distributions while removing temporal ordering. The resulting exponents increase substantially and approach the classical prediction, indicating that heterogeneity alone is insufficient to explain the empirical scaling. 
In contrast, the original careers exhibit truncated power law gap distributions with exponents in the range $0.799 \leq \alpha \leq 0.843$. Since the bootstrap procedure removes only temporal ordering, the difference between the real and shuffled data demonstrates that the sequence of performances within a career plays an important role in record formation. The bootstrap analysis suggests that the observed deviation from the classical 1/g law is primarily associated with the temporal organization of innings within a career. Processes such as improvement, aging, adaptation, and changing roles may introduce long-term temporal structure that alters record statistics.

Reversed careers also yield exponents below unity, showing that the observed behavior cannot be attributed solely to early-career improvement. Overall, our results indicate that personal best progression contains information beyond the underlying score distribution and reflects nontrivial temporal organization within cricket careers.

\section{Conclusion}

We investigated the statistics of inter-record gaps between successive personal best performances in Test, ODI, and T20 cricket. The empirical gap distributions are broad and are well described by a truncated power law. Across all three formats, the fitted scaling exponents lie in the range $(0.799\leq\alpha\leq0.843)$, substantially below the classical prediction $(\alpha=1)$ for independent record processes.

A key result of this work comes from bootstrap analysis. Randomly shuffling the innings of each player preserves both career length and the complete distribution of scores while removing temporal ordering.
After shuffling, the fitted exponents increase substantially, reaching $\alpha$=0.952 for Test cricket, $\alpha$=0.979 for ODI cricket, and $\alpha$=0.939 for T20 cricket. Thus randomizing the temporal order moves the system toward the classical record statistics prediction ($\alpha$=1), although the agreement is not exact. This indicates that temporal ordering contributes significantly to the observed scaling, while additional factors may also influence the record dynamics. None of the synthetic null models reproduces the small empirical exponents, suggesting that the observed statistics arise from more complex career dynamics than captured by these simplified constructions. This demonstrates that the empirical scaling cannot be attributed to differences in player ability, score distributions, or career lengths. Instead, it arises from temporal organization within careers.

 Another point to note is that classical record theory is based on independent continuous random variables where ties have zero probability and the support is essentially unbounded. However, cricket scores are discrete and of finite support. This is an especially important effect in shorter formats. Scores vary a great deal in Test cricket and very large innings are possible, so that the continuous approximation is reasonably accurate. Nevertheless, the difference between the empirical and shuffled exponents that remains is suggestive that temporal organization of performances remains an important contributor to the observed scaling behavior.

The results show that cricket performances are not well represented as independent draws from a stationary distribution. Record progression preserves information about the ordering of performances, thus providing clues about long-term career dynamics such as improvement, adaptation, ageing or changing playing conditions. These effects lead to record statistics which systematically deviate from those of classical record theory.

More broadly, this study shows that personal best progression provides a useful lens through which to examine long-term structure in sporting performance. The observed deviations from classical record statistics highlight the value of record based measures for uncovering temporal organization in complex empirical systems. It would be interesting to determine whether similar departures from classical record statistics occur in other human-performance and non-stationary systems.

\subsection*{Acknowledgments}
PMG thanks IMSc, Chennai, for hosting a visit and Prof. Sitabhra 
Sinha for discussions. We thank Mr. Abhinav Viswaroop for help in 
programming.

\subsection*{Funding}
PDB thanks Rashtrasant Tukadoji Maharaj 
Nagpur University for providing financial assistance (RTMNU/RDC/2024/242).

\subsection*{Declaration of generative AI and AI-assisted technologies}
During the preparation of this work, the authors used chatgpt to improve the language and clarity of this manuscript. The authors reviewed and edited the output as needed and take full responsibility for the content of the published article. 
\subsection*{Conflict of Interest}
The authors declare that they have no known competing financial interests or personal relationships that could have appeared to influence the work reported in this paper.

\subsection*{Data Availability Statement}
The data used in this study were obtained from publicly accessible records available on ESPN Cricinfo. The processed datasets used for the analysis are available from the corresponding author upon reasonable request.


\begin{thebibliography}{99}

\bibitem{boccia2021performance}
Boccia, G., Cardinale, M., and Brustio, P. R. (2021).
Performance progression of elite jumpers: Early performances do not predict later success.
\textit{Scandinavian Journal of Medicine \& Science in Sports}, \textbf{31}(1), 132--139.

\bibitem{gembris2007evolution}
Gembris, D., Taylor, J. G., and Suter, D. (2007).
Evolution of athletic records: Statistical effects versus real improvements.
\textit{Journal of Applied Statistics}, \textbf{34}(5), 529--545.

\bibitem{stevenson2018modelling}
Stevenson, O. G. and Brewer, B. J. (2018).
Modelling career trajectories of cricket players using Gaussian processes.
In \textit{International Conference on Bayesian Statistics in Action}, pp. 165--173. Springer.


\bibitem{berthelot2015has}
Berthelot, G., Sedeaud, A., Marck, A., Antero-Jacquemin, J., Schipman, J., Sauliere, G., Marc, A., Desgorces, F.-D., and Toussaint, J.-F. (2015).
Has athletic performance reached its peak?
\textit{Sports Medicine}, \textbf{45}(9), 1263--1271.

\bibitem{ram2022significant}
Ram, S. K., Nandan, S., and Sornette, D. (2022).
Significant hot hand effect in the game of cricket.
\textit{Scientific Reports}, \textbf{12}(1), 11663.

\bibitem{arnoldrecords}
Arnold, B. C., Balakrishnan, N., and Nagaraja, H. N. (2011).
\textit{Records}. John Wiley \& Sons.

\bibitem{krugrecords}
Krug, J. (2007). Records in a changing world.
\textit{Journal of Statistical Mechanics: Theory and Experiment}, P07001.

\bibitem{Wergenrecords}
Wergen, G., Majumdar, S. N., and Schehr, G. (2012).
Record statistics for multiple random walks.
\textit{Physical Review E}, \textbf{86}, 011119.

\bibitem{ribeirocricket}
Ribeiro, H. V., Costa, L. da F., Rodrigues, F. A., and Andrade Jr., J. S. (2012).
Anomalous diffusion and long-range correlations in cricket scores.
\textit{Physical Review E}, \textbf{86}, 026110.

\bibitem{burstcricket}
Sadekar, O., Chowdhary, S., Santhanam, M. S., and Battiston, F. (2024).
Individual and team performance in cricket.
\textit{Royal Society Open Science}, \textbf{11}(7), 240809.

\bibitem{barabasibursts}
Barab'asi, A.-L. (2005).
The origin of bursts and heavy tails in human dynamics.
\textit{Nature}, \textbf{435}, 207--211.

\bibitem{cricinfostatsguru}
ESPN Cricinfo Statsguru: Cricket Statistics Database(2026).
Available at: \url{https://stats.espncricinfo.com/ci/engine/stats/index.html}
(Accessed 20 April 2026).

\end{thebibliography}

\end{document}